\documentclass[twocolumn,english]{revtex4-1}
\usepackage[T1]{fontenc}
\usepackage[latin9]{inputenc}
\usepackage{amsmath}
\usepackage{amssymb}
\usepackage{graphicx}
\usepackage{babel}
\begin{document}
\title{
\global\long\def\cm{\text{cm}^{-1}}%
Modeling Molecular J and H Aggregates using Multiple-Davydov D2 Ansatz}
\author{Mantas Jaku\v{c}ionis\textsuperscript{1}, Agnius Žukas\textsuperscript{1},
Darius Abramavi\v{c}ius\textsuperscript{1}}
\affiliation{\textsuperscript{1}Institute of Chemical Physics, Vilnius University,
Sauletekio Ave. 9-III, LT-10222 Vilnius, Lithuania}
\begin{abstract}
The linear absorption spectrum of J and H molecular aggregates is
studied using the time-dependent Dirac-Frenkel variational principle
(TDVP) with the multi-Davydov $\text{D}_{2}$ ($\text{mD}_{2}$) trial
wavefunction (Ansatz). Both the electronic and vibrational molecular
degrees of freedom (DOF) are considered. By inspecting and comparing
absorption spectrum of both open and closed chain aggregates over
a range of electrostatic nearest neighbor coupling and temperature
values, we find the $\text{mD}_{2}$ Ansatz to be necessary for obtaining
accurate aggregate absorption spectrum in all parameter regimes considered,
while the regular Davydov $\text{D}_{2}$ Ansatz is not sufficient.
Establishing relation between the model parameters and the depth of
the $\text{mD}_{2}$ Ansatz is the main focus of the study. Molecular
aggregate wavepacket dynamics, during excitation by an external field,
is also studied. We find the wavepacket to exhibit an out-of-phase
oscillatory behavior along the coordinate and momentum axes and an
overall wavepacket broadening, implying the electron-vibrational (vibronic)
eigenstates of an aggregate to reside on non-parabolic energy surfaces.
\end{abstract}
\maketitle

\section{Introduction}

Molecular aggregate excitation dynamics can be computed using the
wavefunction-based TDVP by postulating an Ansatz, which ought to be
complex enough to represent the necessary vibronic states of the aggregate.
The Davydov $\text{D}_{2}$ Ansatz, which was originally developed
for the molecular chain soliton theory \citep{Davydov1979,Scott1991},
represents quantum states of molecular vibrational modes using Gaussian
wavepackets, also known as coherent states (CS). It has been widely
applied to study excitation relaxation processes in both isolated
molecules and in molecular aggregates \citep{Sun2010b,Chorosajev2016c,Jakucionis2018b,Jakucionis2020a},
as well as to compute their linear and nonlinear spectra \citep{Sun2015a,Zhou2016a,Chorosajev2017b,Jakucionis2022}.

While TDVP method is based on propagating pure wavefunctions, its
stochastic extension can be used to describe non-zero temperature
by averaging over initial equilibrium thermal state \citep{Glauber1963}.
However, it still does not properly account for the energy dissipation
effect in the vibronic system. That can be achieved using the thermalization
approach by implicitly modeling vibrational energy exchange with an
extended environment \citep{Jakucionis2021}.

The $\text{D}_{2}$ Ansatz is not sufficient to allow for accurate
modeling of molecular aggregates \citep{Zhou2016}. Accuracy can be
greatly improved by considering a superposition of multiple copies
of the $\text{D}_{2}$ Ansatze, termed the multi-Davydov $\text{D}_{2}$
Ansatz. The $\text{mD}_{2}$ Ansatz, and its more complex variant,
$\text{mD}_{1}$ Ansatz \citep{Zhou2014}, have been applied to study
polaron dynamics in Holstein molecular crystals \citep{Zhou2016},
the spin-boson models \citep{Wang2016} and for nonadiabatic dynamics
of single molecules \citep{Chen2019a,Jakucionis2020a}, as well as
to simulate nonlinear response function of molecular aggregates \citep{Sun2015a,Zhou2016}
and others \citep{Gao2021,Wang2021a,Sun2021,Sun2022}. A more in-depth
overview of the various types of Davydov Ansatze and their applications
can be found in a recent review article by Zhao et al. \citep{Zhao2021}.

However, a well defined strategy to determine the required number
of multiples in $\text{mD}_{2}$ Ansatz (\textit{\emph{ot }}\textit{the
depth) }needed to obtain the converged result is lacking. Absorption
spectrum and excitation relaxation dynamics of a linear molecular
aggregate are key quantities that may serve for establishing relation
between model parameters and the parameters of the Ansatz. Molecule
electronic properties significantly depend on the transition dipoles,
whether the dipoles are in the ``head-to-tail'' (J aggregate) or
``side-to-side'' (H aggregate) configurations \citep{McRae1958,KASHA1963,Kasha1965b,Spano2009,Schroter2015a,Hestand2018}.
In a J aggregate, excitation by an external electric field produces
initially excited lowest energy excitonic state, therefore, energy
relaxation effect is minimal and the absorption spectrum is dominated
by the exchange narrowing effect \citep{Eisfeld2006,Walczak2008a,Roden2008}.
It effectively reduces electron-vibrational coupling strength and
the shape of the spectrum is similar to that of a single molecule,
rescaled due to exchange narrowing. Meanwhile, in an H aggregate,
external fields excite the highest energy excitonic state, thus, various
available vibronic energy relaxation pathways make H aggregate spectra
more complicated than that of the J aggregate, with non-trivial spectral
lineshape \citep{Eisfeld2006,Roden2008}.

The rest of the paper is organized in the following way. First, in
Section \ref{sec:Theory} we describe the vibronic molecular aggregate
model and the theory of linear absorption using the $\text{mD}_{2}$
Ansatz. Secondly, in Section \ref{sec:Results} we analyze a range
of J and H molecular aggregate absorption spectra and quantify their
convergence in terms of $\text{mD}_{2}$ Ansatz depth. Lastly, in
Section \ref{sec:Discussion}, we discuss our findings, relate $\text{mD}_{2}$
Ansatz vibrational wavepacket evolution to the previously proposed
$\text{sD}_{2}$ Ansatz and present conclusions.

\section{Theory and its numerical implementation\label{sec:Theory}}

We consider a vibronic molecular aggregate model, where both the electronic
and the vibrational DOF are included. Each molecule (\textit{site})
in the aggregate is modeled as a two electronic-level system, where
$\varepsilon_{n}$ is the $n$th site excited electronic state energy.
Electrostatic interaction between excited electronic states of sites
is given in terms of the resonant dipole-dipole interaction with strengths
$J_{nm}$. Intramolecular vibrational modes of sites are modeled as
harmonic vibrational modes. Mode $q$ of the $k$th site is characterized
by a frequency $w_{kq}$ and the electron-vibrational coupling strength
$f_{kq}$.

Vibronic aggregate model Hamiltonian $\hat{H}$ is given as a sum
of the following Hamiltonians \citep{Valkunas2013a,Bardeen2014,Amerongen2010,Schroter2015}.
Electronic site Hamiltonian 
\begin{equation}
\hat{H}_{\text{S}}=\sum_{n}\varepsilon_{n}\hat{a}_{n}^{\dagger}\hat{a}_{n}+\sum_{n,m}^{n\neq m}J_{nm}\hat{a}_{n}^{\dagger}\hat{a}_{m},
\end{equation}
 describes an electronic excitation delocalized over the whole aggregate
(\textit{exciton}), where $\hat{a}_{n}^{\dagger}$ $\left(\hat{a}_{n}\right)$
are the $n$th site Paulionic excitation creation (annihilation) operators.
Intramolecular vibrational mode Hamiltonian (with the reduced Planck's
constant set to $\hbar=1$) is that of quantum harmonic oscillators
(QHO) 
\begin{equation}
\hat{H}_{\text{V}}=\sum_{k,q}w_{kq}\hat{c}_{kq}^{\dagger}\hat{c}_{kq},
\end{equation}
with excluded zero-quanta energy constant shift, where $\hat{c}_{kq}^{\dagger}$
$\left(\hat{c}_{kq}\right)$ are oscillator bosonic creation (annihilation)
operators of the $q$th intramolecular mode, coupled to the $k$th
site, which account for molecular vibrations. The electronic-vibrational
interaction is included using the shifted oscillator model, i.e.,
the vibrational mode potential becomes displaced along the coordinate
axis in the excited electronic state. Electron-vibrational coupling
Hamiltonian is then given by 
\begin{align}
\hat{H}_{\text{S-V}}= & -\sum_{n}\hat{a}_{n}^{\dagger}\hat{a}_{n}\sum_{q}w_{nq}f_{nq}\left(\hat{c}_{nq}^{\dagger}+\hat{c}_{nq}\right).
\end{align}

Molecular aggregate sites also interact with an external electric
field $\boldsymbol{E}\left(t\right)=\boldsymbol{e}E\left(t\right)\exp\left(-\text{i}\omega_{\text{field}}t\right)$,
where $\boldsymbol{e}$ is the optical polarization vector, $E\left(t\right)$
is the time-dependent field envelope and $\omega_{\text{field}}$
is the field frequency. In the dipole and Frank-Condon approximations,
sites interact with optical electric field via their purely electronic
transition dipole vectors $\boldsymbol{\mu}_{n}$, therefore, the
site-field coupling Hamiltonian is given as $\hat{H}_{\text{S-F}}\left(t\right)=\hat{\boldsymbol{\mu}}\cdot\boldsymbol{E}\left(t\right)$
with $\hat{\boldsymbol{\mu}}=\hat{\boldsymbol{\mu}}_{+}+\hat{\boldsymbol{\mu}}_{-}$
being the transition dipole operator and
\begin{align}
\hat{\boldsymbol{\mu}}_{+} & =\sum_{n}\boldsymbol{\mu}_{n}\hat{a}_{n}^{\dagger},\label{eq:dipUP}\\
\hat{\boldsymbol{\mu}}_{-} & =\sum_{n}\boldsymbol{\mu}_{n}\hat{a}_{n},\label{eq:dipDOWN}
\end{align}
are the transition operators that increase (decrease) the number of
excitation quanta in the aggregate. We consider electric field in
an impulsive limit with rotating wave approximation \citep{Mukamel1995a},
$E\left(t\right)\rightarrow E_{0}\delta\left(t-\tau\right)$, where
$\tau$ is the interaction time, therefore, transitions between aggregate
states with different number of excitations occur instantaneously.

Using the Heitler-London approach \citep{Frenkel1931,Valkunas2013a},
we construct the electronic states of the aggregate as products of
molecular excitations: the molecular aggregate electronic ground state
$|0\rangle=\bigotimes_{n}|0_{n}\rangle$ (global ground state of all
sites) is taken as a reference state, thus, in the ground state, inter-site
coupling and electron-vibrational coupling are absent, we also have
the electronic ground state energies equal to zero. Then the aggregate
ground state Hamiltonian is purely vibrational $\hat{H}_{G}=\hat{H}_{\text{V}}$.

Time propagation of various states be computed using TDVP applied
to the Davydov Ansatze \citep{Sun2010b,Zhou2016,Jakucionis2018b}.
Since the \emph{ground} electronic state $\left(\text{g}\right)$
corresponds to independent molecular vibrations, it is sufficient
to describe\textit{ }it by the simplest $\text{D}_{2}$ Ansatz
\begin{equation}
|\Psi_{\text{D}_{2}}^{\left(\text{g}\right)}\left(t\right)\rangle=\vartheta\left(t\right)|0\rangle\otimes|\boldsymbol{\lambda}\left(t\right)\rangle,\label{eq:PSI_0}
\end{equation}
where $\vartheta\left(t\right)$ is the ground state amplitude. Vibrational
state is represented in terms of the multi-dimensional CS, $|\boldsymbol{\lambda}\left(t\right)\rangle=\bigotimes_{k,q}|\lambda_{kq}\left(t\right)\rangle$.
Single-dimensional CS $|\lambda_{kq}\left(t\right)\rangle$ is created
by applying translation operator 
\begin{equation}
\hat{D}\left(\lambda_{kq}\left(t\right)\right)=\exp\left(\lambda_{kq}\left(t\right)\hat{c}_{kq}^{\dagger}-\lambda_{kq}^{\star}\left(t\right)\hat{c}_{kq}\right),
\end{equation}
with complex displacement parameter $\lambda_{kq}\left(t\right)$,
to the QHO vacuum state: $\hat{D}\left(\lambda_{kq}\left(t\right)\right)|0\rangle_{kq}=|\lambda_{kq}\left(t\right)\rangle$.
For the time propagation of the aggregate's electronic \emph{excited}
state $\left(\text{e}\right)$, $\text{mD}_{2}$ Ansatz will be used
\citep{Zhou2016}, given by
\begin{equation}
|\Psi_{\text{mD}_{2}}^{\left(\text{e}\right)}\left(t\right)\rangle=\sum_{i}^{M}\sum_{n}\alpha_{i,n}\left(t\right)|n\rangle\otimes|\boldsymbol{\lambda}_{i}\left(t\right)\rangle,\label{eq:mD2}
\end{equation}
where $|n\rangle=|1_{n}\rangle\bigotimes_{m\neq n}|0_{m}\rangle$
is an electronic state of amplitude $\alpha_{i,n}\left(t\right)$,
which defines a singly excited $n$th site. Aggregate's vibrational
state now is $|\boldsymbol{\lambda}_{i}\left(t\right)\rangle=\bigotimes_{k,q}|\lambda_{i,kq}\left(t\right)\rangle$.
Each multiple $i$ corresponds to an excitonic state associated with
an aggregate vibrational state. By considering more multiples, complexity
and, in principle, accuracy of the $\Psi_{\text{mD}_{2}}^{\left(\text{e}\right)}$
could be increased. The $\Psi_{\text{mD}_{2}}^{\left(\text{e}\right)}$
Ansatz with $M=1$ reduces to the regular Davydov $\Psi_{\text{D}_{2}}^{\left(\text{e}\right)}$
Ansatz.

While, in general, the state of the aggregate is the superposition
of the ground $\Psi_{\text{D}_{2}}^{\left(\text{g}\right)}$ and the
excited $\Psi_{\text{mD}_{2}}^{\left(\text{e}\right)}$ state wavefunctions,
in the perturbative treatment of interaction with the optical field,
the aggregate's electronic state will always adhere to either $\Psi_{\text{D}_{2}}^{\left(\text{g}\right)}$
or $\Psi_{\text{mD}_{2}}^{\left(\text{e}\right)}$, therefore it is
sufficient to consider evolution of these wavefunctions independently.

For the ground state, TDVP procedure results in a system of explicit
differential equations of motion (EOM) for variables $\vartheta\left(t\right)$,
$\lambda_{kq}\left(t\right)$, which yield analytical solution: $\vartheta\left(t\right)=\vartheta\left(0\right)$,
$\lambda_{kq}\left(t\right)=\exp\left(-\text{i}w_{kq}t\right)$, while
for the electronic excited state, the resulting EOM constitute a system
of implicit differential equations for $\alpha_{i,n}\left(t\right)$,
$\lambda_{i,kq}\left(t\right)$ variables, which can be solved numerically.
Details on the $\text{mD}_{2}$ Ansatz EOM, their solution and numerical
implementation, can be found in Appendix \ref{sec:Appendix-A}.

Using the response function theory \citep{Mukamel1995a,Valkunas2013a},
the linear absorption spectrum is given by a half-Fourier transform,
\begin{equation}
A\left(\omega\right)=\text{Re}\int_{0}^{\infty}\text{d}t\text{e}^{i\omega t-\gamma t}S^{\left(1\right)}\left(t\right),\label{eq:Abs}
\end{equation}
 of the linear response function $S^{\left(1\right)}\left(t\right)$,
given by
\begin{equation}
S^{\left(1\right)}\left(t\right)=\langle\Psi_{\text{D}_{2}}^{\left(\text{g}\right)}\left(0\right)|\hat{\mu}_{-}e^{i\hat{H}t}\hat{\mu}_{+}e^{-i\hat{H}_{\text{G}}t}|\Psi_{\text{D}_{2}}^{\left(\text{g}\right)}\left(0\right)\rangle,\label{eq:S1}
\end{equation}
where we defined a scalar dipole operator $\hat{\mu}_{\pm}=\boldsymbol{e}\cdot\hat{\boldsymbol{\mu}}_{\pm}$.
We also include phenomenological dephasing rate of $\gamma=100\ \cm$
to account for the decay of optical coherence due to the environment
fluctuations, explicitly unaccounted by our approach.

Numerical computation of $S^{\left(1\right)}\left(t\right)$ can be
greatly streamlined by deriving expressions that relate variables
of the ground state $\vartheta\left(t\right)$, $\lambda_{kq}\left(t\right)$
and the excited state $\alpha_{i,n}\left(t\right)$, $\lambda_{i,kq}\left(t\right)$,
when an upward transition operators $\hat{\mu}_{+}$ act on the ground
state $\Psi_{\text{D}_{2}}^{\left(\text{g}\right)}$, such that we
can define
\begin{align}
\hat{\mu}_{+}|\Psi_{\text{D}_{2}}^{\left(\text{g}\right)}\left(\tau\right)\rangle\equiv & |\Psi_{\text{mD}_{2}}^{\left(\text{e}\right)}\left(\tau\right)\rangle,\label{eq:0to1}
\end{align}
and, from Eqs. (\ref{eq:dipUP}), (\ref{eq:dipDOWN}), follows also
that 
\begin{equation}
\langle\Psi_{\text{D}_{2}}^{\left(\text{g}\right)}\left(\tau\right)|\hat{\mu}_{-}\equiv\left(\hat{\mu}_{+}|\Psi_{\text{D}_{2}}^{\left(\text{g}\right)}\left(\tau\right)\rangle\right)^{\dagger}.\label{eq:eq0to1_cj}
\end{equation}
Notice, that, in general, even though at the time of interaction $\tau$,
initial state $\Psi_{\text{D}_{2}}^{\left(\text{g}\right)}$ is normalized,
the resulting excited state $\Psi_{\text{mD}_{2}}^{\left(\text{e}\right)}$
is not necessarily normalized. This does not introduce difficulties,
since in the derivation of EOM, no assumptions of Ansatze normalization
has been made. Alternatively, the resulting wavefunctions from Eqs.
(\ref{eq:0to1}), (\ref{eq:eq0to1_cj}) can be manually normalized,
however, this would require keeping track of excitation amplitudes
separately.

During the ground to the excited state transition in Eq. (\ref{eq:0to1}),
the ground state wavefunction $\Psi_{\text{D}_{2}}^{\left(\text{g}\right)}$
can be equivalently represented by an arbitrary single CS out of the
$i=1,2,\ldots,M$ multiples of the $\Psi_{\text{mD}_{2}}^{\left(\text{e}\right)}$
Ansatz. For this reason, we choose to ``populate'' the $i=1$ multiple
after excitation, and call the rest of the multiples $j\neq i$ as
initially ``unpopulated''. Then the newly created state $\Psi_{\text{mD}_{2}}^{\left(\text{e}\right)}$,
given by Eq. (\ref{eq:0to1}), has amplitudes $\alpha_{i=1,n}\left(\tau\right)=\mu_{n}\vartheta\left(\tau\right)$,
$\alpha_{j,n}\left(\tau\right)=0$, where $\mu_{n}=\boldsymbol{e}\cdot\boldsymbol{\mu}_{n}$,
and CS displacements $\lambda_{i=1,kq}\left(\tau\right)=\lambda_{kq}\left(\tau\right)$.

Unpopulated CS variables $\lambda_{j,kq}\left(\tau\right)$ initially
do not contribute to the dynamics, therefore, their position, in principle,
is arbitrary. However, during the following excited state evolution,
unpopulated multiples become populated and begin to influence model
dynamics. It is known, that the initial distance between the populated
and unpopulated CS $\delta=\left|\lambda_{i=1,kq}\left(\tau\right)-\lambda_{j,kq}\left(\tau\right)\right|$
should not be too large, otherwise, they will not participate in the
excited state dynamics (even at large propagation times CS will remain
separated) \citep{Jakucionis2020a}. On the other hand, setting all
CS in close proximity to each other $\lambda_{j,kq}\left(\tau\right)\approx\lambda_{i=1,kq}\left(\tau\right)$,
leads to a highly singular EOM \citep{Bargmann1971,Werther2020}.
We chose to set unpopulated CS in a layered hexagonal pattern around
the populated CS given by equation
\begin{align}
\lambda_{j,kq}\left(\tau\right) & =\lambda_{i=1,kq}\left(\tau\right)\nonumber \\
 & +\Delta\sin\left(\frac{\pi}{m}\right)\left(1+\left\lfloor \beta\right\rfloor \right)e^{\text{i}2\pi\left(\beta+\frac{1}{2m}\left\lfloor \beta\right\rfloor \right)}
\end{align}
where $\Delta$ is a distance parameter, $\beta\left(j,m\right)=\left(j-2\right)/m$
is a coordination function with $m=6$ being the number of CS in each
layer and $\left\lfloor x\right\rfloor $ is the floor function of
$x$. $\Delta$ should be large enough not to have significant overlap
among the initial distribution of CS, we found $\Delta=0.5$ to give
numerically well behaved, consistent and convergent results.

After independently propagating \textit{bra} $\left(\text{L}\right)$
and \textit{ket} $\left(\text{R}\right)$ states of Eq. (\ref{eq:S1}),
their overlap is given by
\begin{align}
S^{\left(1\right)}\left(t\right) & =\langle\Psi_{\text{mD}_{2}}^{\left(\text{e}\right)}\left(t\right)|_{\text{L}}\cdot|\Psi_{\text{mD}_{2}}^{\left(\text{e}\right)}\left(t\right)\rangle_{\text{R}}\nonumber \\
 & =\sum_{i,j}\sum_{n}\alpha_{i,n}^{\star\left(\text{L}\right)}\left(t\right)\alpha_{j,n}^{\left(\text{R}\right)}\left(t\right)\langle\boldsymbol{\lambda}_{i}\left(t\right)|_{\text{L}}\cdot|\boldsymbol{\lambda}_{j}\left(t\right)\rangle_{\text{R}},
\end{align}
where the CS overlap is given by 
\begin{align}
\langle\boldsymbol{\lambda}_{i}\left(t\right)|_{\text{L}}\cdot|\boldsymbol{\lambda}_{j}\left(t\right)\rangle_{\text{R}} & =\exp\sum_{k,q}\left(\lambda_{i,kq}^{\star\left(\text{L}\right)}\left(t\right)\lambda_{j,kq}^{\left(\text{R}\right)}\left(t\right)\right)\nonumber \\
 & \times\exp\sum_{k,q}\left(-\frac{1}{2}\left|\lambda_{i,kq}^{\star\left(\text{L}\right)}\left(t\right)\right|^{2}\right)\nonumber \\
 & \times\exp\sum_{k,q}\left(-\frac{1}{2}\left|\lambda_{j,kq}^{\left(\text{R}\right)}\left(t\right)\right|^{2}\right).
\end{align}

Temperature of the molecular aggregate is included by implementing
the Monte Carlo ensemble averaging scheme. Before excitation of molecular
aggregate via an external field, vibrational modes reside in the ground
state and obey the canonical ensemble statistics with density operator
in the $P$-representation given by the probability function \citep{Glauber1963,Chorosajev2016c,Wang2017b,Xie2017}
\begin{equation}
\mathcal{P}\left(\lambda_{kq}\left(0\right)\right)=\mathcal{Z}_{kq}^{-1}\exp\left(-\left|\lambda_{kq}\left(0\right)\right|^{2}\left[\text{e}^{\frac{\omega_{kq}}{k_{\text{B}}T}}-1\right]\right),\label{eq:Sudarshan-P}
\end{equation}
where $\mathcal{Z}_{kq}$ is the partition function, $k_{\text{B}}$
is the Boltzmann constant and $T$ is the temperature. By sampling
vibrational mode initial conditions $\lambda_{kq}\left(0\right)$
from Eq. (\ref{eq:Sudarshan-P}), and averaging over the linear response
functions $S^{\left(1\right)}\left(t\right)$, we obtain thermally
averaged linear response function $\left\langle S^{\left(1\right)}\left(t\right)\right\rangle _{T}$,
which now depends on the temperature. We found $360$ samples to result
in converged absorption spectrum presented in the next Section.

\section{Results\label{sec:Results}}

We consider the absorption spectra of the H and J aggregates. The
model aggregate consists of $N=10$ sites, each of which can be resonantly
excited by an external electric field, thus we set single site excitation
energies to $\varepsilon_{n}=\omega_{\text{field}}$ with the nearest
neighbor couplings $J_{n,n+1}=J=\pm500\ \cm$ for H and J aggregates,
respectively. For each aggregate type, we consider two types of boundary
conditions: open chain (OC) with $J_{N,1}=J_{1,N}=0\ \cm$, and the
closed chain (CC) with $J_{N,1}=J_{1,N}=J$. Purely excitonic absorption
spectrum of such CC aggregate consists of a single peak due to the
superradiant excitonic state with $\varepsilon_{n}+2J$ energy. The
OC aggregate, besides having the main peak at $\approx\varepsilon_{n}+2J$,
also has many lower amplitude peaks.

Next, we include one intramolecular vibrational mode per site with
frequency $\omega_{kq}=500\ \cm$ and Huang-Rhys (HR) factor $S=f_{kq}^{2}=1$,
which defines the electron-vibrational coupling strength. Site electronic
transition dipole moment vectors are identical and set to $\boldsymbol{\mu}_{n}=\left(1,0,0\right)$.
Vibrational mode initial thermal energy is set to $k_{\text{B}}T=\omega_{kq}/2$,
which corresponds to the temperature of $T=360\ \text{K}.$ Note,
that rescaling all energy parameters by a constant would give exactly
the same spectrum.

\begin{figure}
\includegraphics[width=8.5cm]{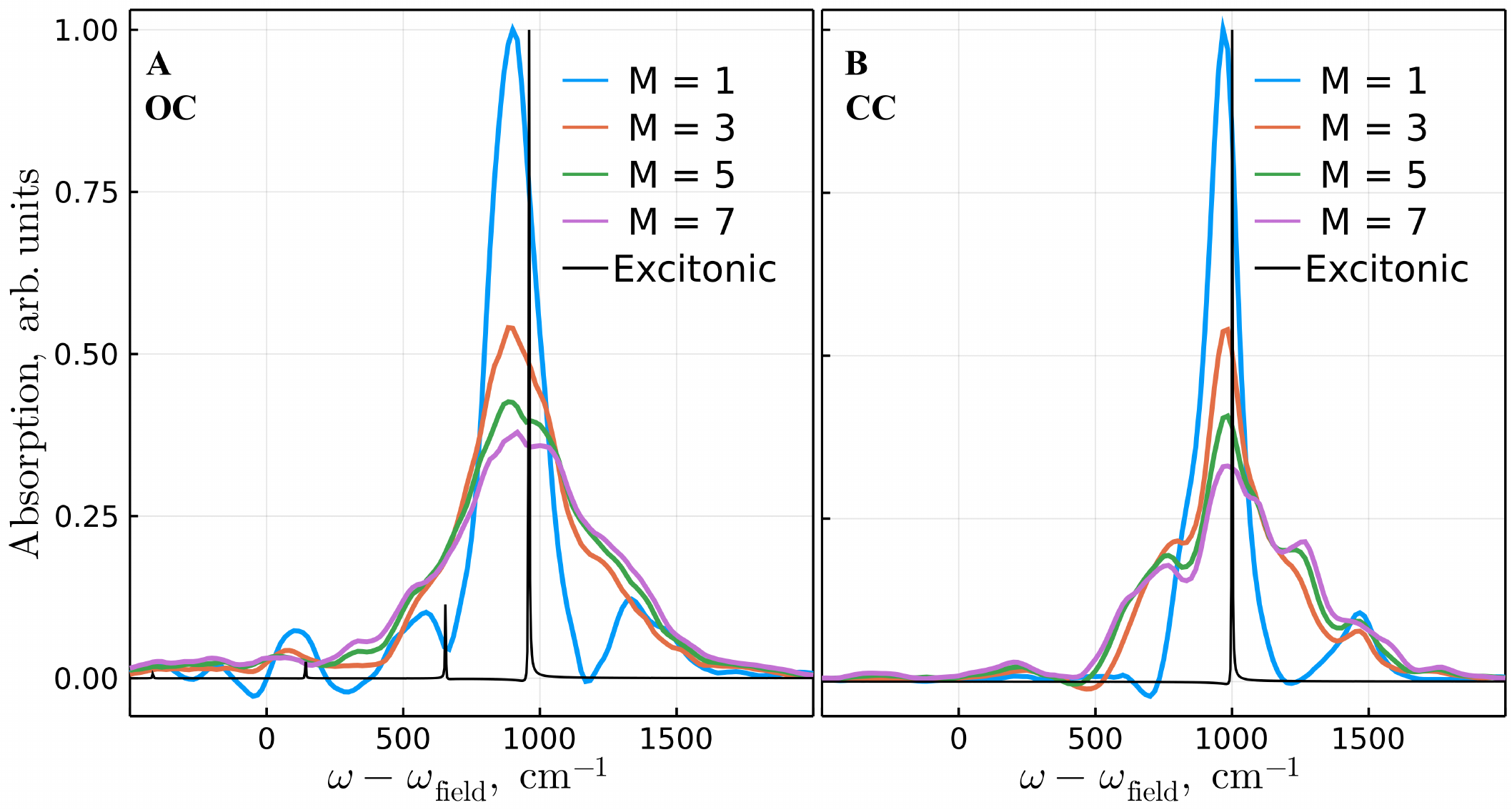}

\caption{Absorption spectrum of model H aggregate in (A) OC and (B) CC configurations,
computed with $\text{mD}_{2}$ Ansatz depth $M$. Purely excitonic
spectra is also shown.\label{fig:H-absorption}}
\end{figure}

Absorption spectrum of the model H aggregate, computed with an increasing
$\text{mD}_{2}$ Ansatz depth $M$, both in OC and CC arrangement,
are shown in Fig. (\ref{fig:H-absorption}). In both cases, absorption
spectrum converges with $M=7$ multiples, higher multiplicity spectra
have been computed and are identical up to $M=11$. Absorption of
the $M=1$ case, which is equivalent to using the Davydov $\text{D}_{2}$
Ansatz, has peaks in the same frequencies as the converged spectrum,
however, their intensities are incorrect, some are even negative.
By increasing the number of multiples considered, peak amplitudes
become strictly positive. The 0-0 electronic peak can be clearly identified.
Vibrational side-peaks to the higher energy side are due to 0-n vibronic
transitions, while on the lower energy side reside the n-0 transition
peaks, permitted by the non-zero temperature. Finite vibronic peak
widths originate from the vibrational dephasing, due to finite temperature
and aggregate environment fluctuations. Both the OC and CC aggregates
have similar lineshapes, slightly finer vibronic structure can be
observed in the CC system, due to a larger symmetry and, therefore,
effectively lower broadening.

\begin{figure}
\includegraphics[width=8.5cm]{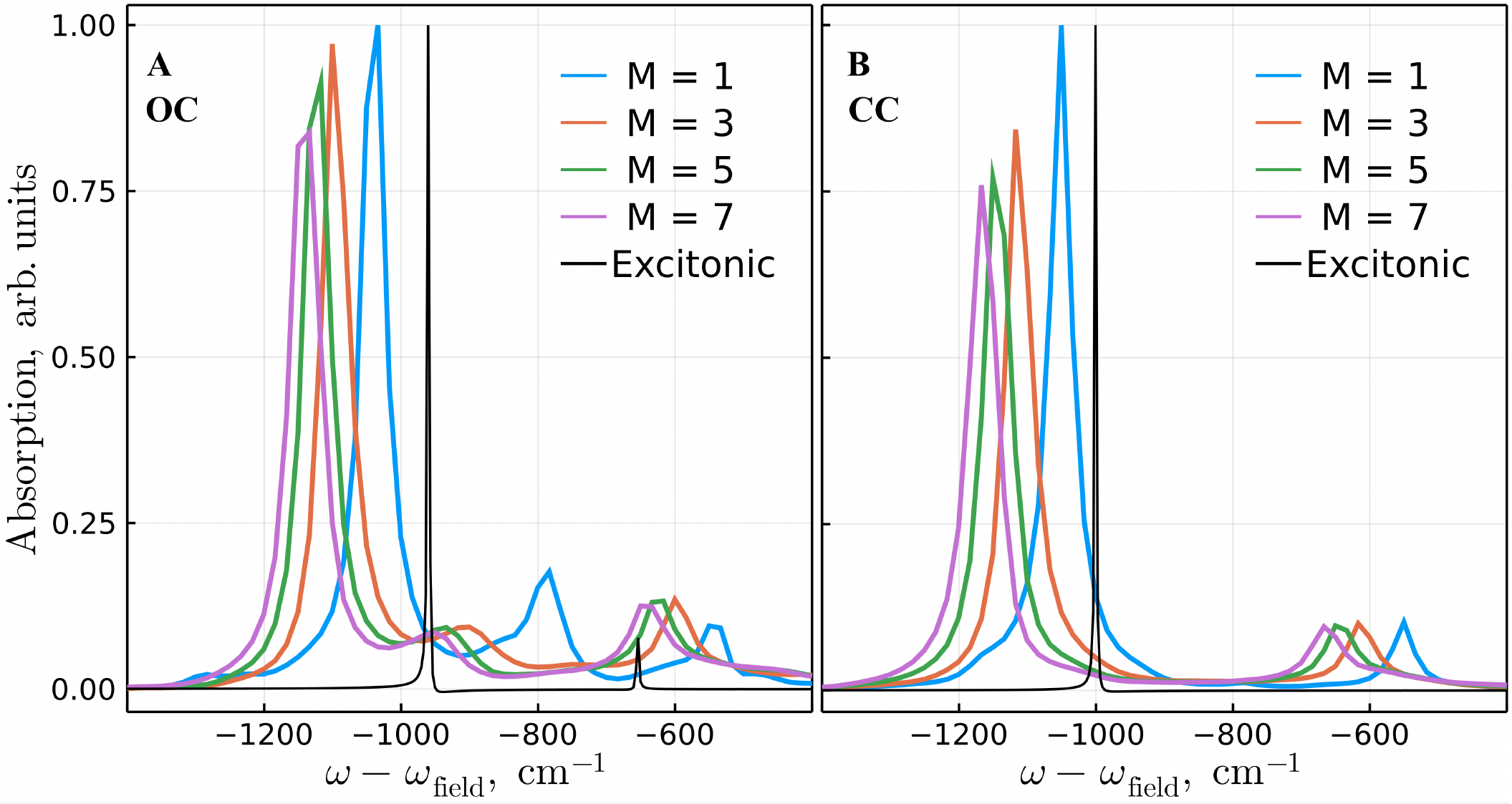}

\caption{Absorption spectrum of model J aggregate in (A) OC and (B) CC configurations,
computed with $\text{mD}_{2}$ Ansatz depth $M$. Purely excitonic
spectra is also shown.\label{fig:J-absorption}}
\end{figure}

Absorption spectrum of the model J aggregate is shown in Fig. (\ref{fig:J-absorption}).
In the CC aggregate, visible side-peak, on the higher energy side
of the strong 0-0 transition, is the first term of vibrational progression.
The effective HR factor is thus significantly reduced (hence, the
exchange narrowing) due to intermolecular couplings. It is observed
independent of Ansatz depth considered in both OC and CC arrangements.
By increasing $M$, absorption spectrum redshifts to lower energies,
while qualitatively maintaining the same shape, however, slight differences
emerge. For the OC aggregate, peak intensities change, while for CC
aggregate, mostly only the main peak intensity changes. Apparent energy
splitting between electronic transitions is considerably reduced,
implying that vibronic states do not maintain excitonic intraband
gaps due to the strong intramolecular vibrational coupling. In contrast
to the H aggregate, all absorption peaks are positive, even with $M=1$
multiplicity.

\begin{figure}
\includegraphics[width=8.5cm]{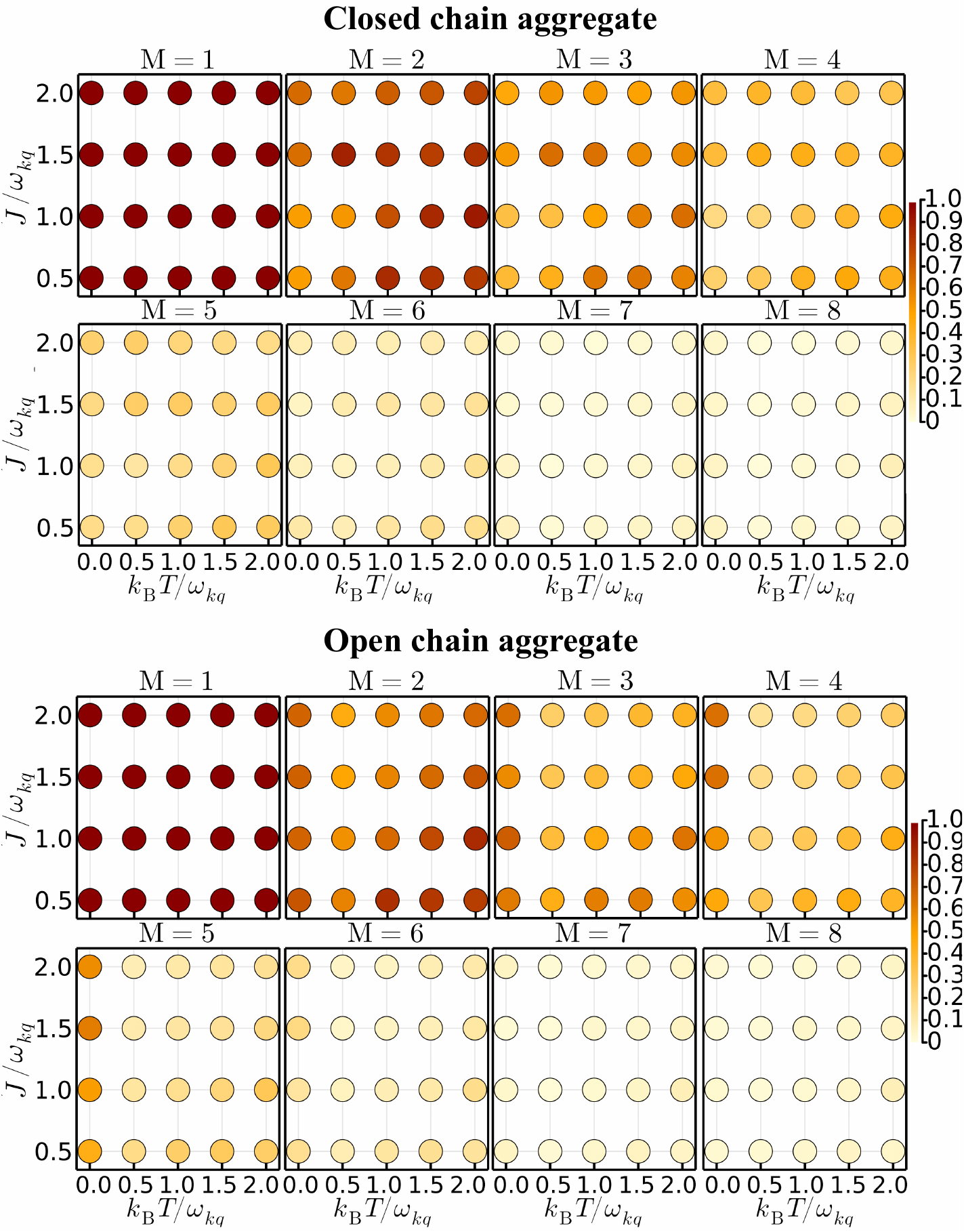}

\caption{Normalized discrepancy $\mathcal{D}\left(M\right)$ of the H aggregate
in a CC and OC configurations for a range of $J$, $k_{\text{B}}T$
values, computed with $\text{mD}_{2}$ Ansatz depth $M$.\label{fig:JH-map}}
\end{figure}

In order to quantify convergence of H aggregate absorption spectrum
with increasing$\text{mD}_{2}$ Ansatz depth, we calculate the normalized
discrepancy \citep{Gelzinis2015} 
\begin{equation}
\mathcal{D}\left(M\right)=\frac{1}{\mathcal{N}}\int\text{d}\omega\sqrt{\left(A\left(\omega,M\right)-\overline{A}\left(\omega\right)\right)^{2}},\label{eq:discprence}
\end{equation}
where $A\left(\omega,M\right)$ is the absorption spectrum with multiplicity
$M$, where $\overline{A}\left(\omega\right)=A\left(\omega,M=11\right)$
is the converged reference spectra and 
\begin{equation}
\mathcal{N}=\max_{\text{over}\ M}\int\text{d}\omega\sqrt{\left(A\left(\omega,M\right)-\overline{A}\left(\omega\right)\right)^{2}},
\end{equation}
is the normalization factor. In Fig. (\ref{fig:JH-map}) we show $\mathcal{D}\left(M\right)$
for H aggregate for various values of the nearest neighbor coupling
$J$, vibrational mode thermal energy $k_{\text{B}}T$, and Ansatz
depth $M=1,\ldots,8$.

We observe that for CC and OC H aggregates, discrepancy significantly
depends on $J$ and $k_{\text{B}}T$, even at the same depth $M$.
We observe, that in the case of $M=1,$ independent of model parameters
and site arrangement, spectrum discrepancy is always high. By increasing
depth to just $M=2$, for some parameters, discrepancy is reduced
significantly. By inspecting higher depths ($M=2-4$), a general observation
can be made. Mainly, that the CC H aggregate requires larger depth
at higher temperatures, while for the OC H aggregate, two parameter
regions of high discrepancy can be discerned: at low temperatures,
independent of the coupling strength, and at high temperatures at
weak coupling. The high temperature cases can be rationalized as needing
more CS to represent thermally excited QHO eigenstates with quantum
numbers $n>0$, which are more probable at higher temperatures. Reasoning
for the low temperature case is more subtle. Aggregate excitation
via an external field shifts oscillators away from their equilibrium
considerably (HR factor $S=1$), then the molecular wavepacket relaxes
via vibronic state energy surfaces, which induce wavepacket shape
changes and/or wavepacket splitting between vibronic surfaces. Either
of these two effects would necessitate molecular aggregate wavepacket
to be represented by the $\text{mD}_{2}$ Ansatz with depth $M>1$.

As seen in Fig. (\ref{fig:J-absorption}), by considering larger depth,
J aggregate absorption spectrum lineshape redshifts with minimal changes
to the overall shape of the spectrum, therefore, the use of discrepancy
estimate in Eq. (\ref{eq:discprence}) is not necessary. By visually
inspecting spectrum of J aggregate with various nearest neighbor couplings
and temperatures (shown in Supplementary Material), we find depth
of $M=7$ to again give a well converged result.

\section{Discussion\label{sec:Discussion}}

The total energy of a single QHO, represented by the $\text{D}_{2}$
Ansatz, is proportional to the CS displacement $\lambda$ from the
origin, $E_{\text{D}_{2}}^{\left(\text{osc}\right)}\propto|\lambda|^{2}$,
and the wavepacket shape is that of the lowest energy QHO eigenstate
with quantum number $n=0$, i.e., a simple Gaussian. On the other
hand, using representation of the $\text{mD}_{2}$ Ansatz, oscillator
energy is proportional to the sum of products of CS displacements,
$E_{\text{mD}_{2}}^{\left(\text{osc}\right)}\propto\sum_{i,j}\lambda_{i}^{\star}\lambda_{j}$,
and the wavepacket now is not necessarily a Gaussian due to the interference
of multiple CS. This allows $\text{mD}_{2}$ Ansatz to represent more
complicated QHO eigenstate wavepackets with quantum numbers $n>0$.
It should be noted, that CS can be used to represent an arbitrary
wavepacket using the unity operator expression
\begin{equation}
\hat{I}=\pi^{-1}\iint\text{d\text{Re}\ensuremath{\lambda}}\ \text{d\text{Im}\ensuremath{\lambda}}\ |\lambda\rangle\langle\lambda|,
\end{equation}
consequently, $\text{mD}_{2}$ Ansatz with infinite depth would allow
for complete and exact description of a quantum system. It thus becomes
important to obtain the lower limit at which the vibronic dynamics
is properly described for e. g. absorption spectroscopy.

\begin{figure}
\includegraphics[width=8.5cm]{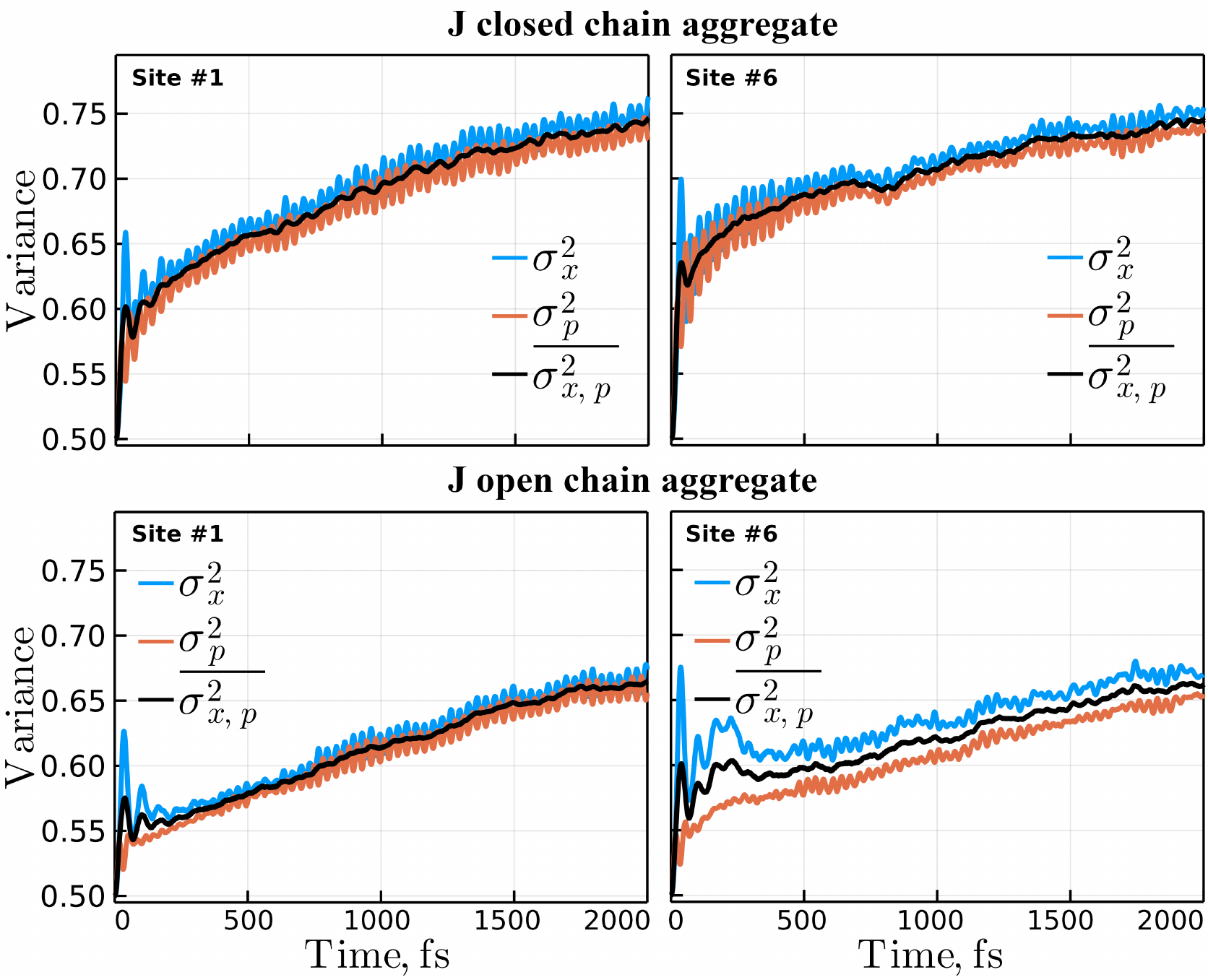}

\caption{Coordinate, momentum and their average variances of a 1st and 6th
site vibrational mode of a J aggregate in ring and chain configurations
with multiplicity $M=11$. \label{fig:variances}}
\end{figure}

Using either of the $\text{D}_{2}$ or $\text{mD}_{2}$ representations,
oscillator can have equal energy, $E_{\text{D}_{2}}^{\left(\text{osc}\right)}=E_{\text{mD}_{2}}^{\left(\text{osc}\right)}$,
however, their wavepacket shape must not be equivalent. It is therefore
interesting to look at vibrational mode wavepacket transition from
being represented by the $\text{D}_{2}$ to a more complex $\text{mD}_{2}$
Ansatz. Such transition occurs naturally in Eq. (\ref{eq:S1}), for
the computation of the linear response function, when an upward transition
dipole operator acts on the aggregate ground state, as given by Eq.
(\ref{eq:0to1}). One way to track wavepacket changes, is to consider
its coordinate and momentum variances, given by
\begin{align}
\sigma_{x}^{2}\left(t\right) & =\left\langle \hat{x}^{2}\left(t\right)\right\rangle -\left\langle \hat{x}\left(t\right)\right\rangle ^{2},\\
\sigma_{p}^{2}\left(t\right) & =\left\langle \hat{p}^{2}\left(t\right)\right\rangle -\left\langle \hat{p}\left(t\right)\right\rangle ^{2},
\end{align}
where $\left\langle \mathcal{O}\left(t\right)\right\rangle =\left\langle \Psi_{\text{mD}_{2}}^{\left(\text{e}\right)}\left(t\right)\left|\hat{\mathcal{O}}\right|\Psi_{\text{mD}_{2}}^{\left(\text{e}\right)}\left(t\right)\right\rangle $
is an expectation value of operator $\hat{\mathcal{O}}$, and their
average variance
\begin{equation}
\overline{\sigma_{x,p}^{2}}\left(t\right)=\frac{1}{2}\left(\sigma_{x}^{2}\left(t\right)+\sigma_{p}^{2}\left(t\right)\right).
\end{equation}
For an independent QHO, the average variance is $\overline{\sigma_{x,p}^{2}}=n+\frac{1}{2}$,
where $n$ is the QHO occupation number. In Fig. (\ref{fig:variances})
we display coordinate, momentum and their average variances of vibrations
coupled to the 1st and 6th sites of the J aggregate in both OC and
CC configurations with depth $M=10$. In the OC configuration, 1st
site is the outermost and the 6th site is in the middle of the aggregate,
while in the CC, these modes are translationally invariant and represent
two modes with a largest separation.

In both configurations, we observe coordinate and momentum variance
oscillations in an out-of-phase manner, while at the same time, the
average variance also increases, slightly more in a CC. Instead of
considering superposition of CS to capture such oscillatory behavior,
squeezed coherent states (SCS) could be used \citep{Tsm1991,Abramavicius2018d,Chorosajev2017c,Zeng2022},
which are able to produce similar variance oscillations intrinsically
. Downside of using SCS would be the need to additionally propagate
variables describing squeezing amplitude and phase for each vibrational
mode. In the Davydov $\text{D}_{2}$ type Ansatz with SCS, the overall
increase of variance would not be captured, yet, for low temperatures,
this might serve as a sufficient approximation. For high temperatures,
multi-Davydov $\text{D}_{2}$ type Ansatz with SCS would be required.

In a CC configuration, due to the 10-fold symmetry of the aggregate,
no difference between variance changes of vibrational modes is to
be expected, while slight differences are observed due to the finite
size of the thermal ensemble considered. Meanwhile, in the OC, difference
between variances of the outer and inner modes can be seen. Outer
vibrational mode, again, show increasing, but oscillatory dynamics,
while the inner mode coordinate and momentum variance values differ,
i.e., wavepacket becomes permanently more stretched along the momentum
axis as compared to the coordinate axis.

We see, that vibrational mode variance changes without any explicit
coupling term between the vibrational DOF. Previously we have proposed
a simplified version of the $\text{mD}_{2}$ Ansatz, termed $\text{sD}_{2}$
\citep{Jakucionis2020a}, by considering multiplicity only of vibrational
mode states. We have observed that the energy transfer between vibrational
modes required inclusion of quadratic or higher order Hamiltonian
coupling term between oscillators, which deformed the initially quadratic
oscillator potential energy surfaces. Energy transfer between vibrational
modes manifested itself as an increase of vibrational mode variance.
In the presented case of the $\text{mD}_{2}$ Ansatz, vibrational
mode variance increase without introducing any explicit Hamiltonian
coupling terms, implying that the multiplicity of the vibronic states
implicitly changes the parabolic potential energy surfaces into non-parabolic.
This can be understood by solving for vibronic energy surfaces the
eigenstates $E(x_{1},\ldots,x_{Q})$. E.g., for a dimer aggregate,
vibronic aggregate Hamiltonian $\hat{H}$ characteristic polynomial
equation is equal to
\begin{align}
0 & =\left(\varepsilon_{1}+\omega f^{2}-x_{1}\omega f+\frac{\omega}{2}\left(x_{1}^{2}+x_{2}^{2}\right)-E(x_{1},x_{2})\right)\nonumber \\
 & \times\left(\varepsilon_{2}+\omega f^{2}-x_{2}\omega f+\frac{\omega}{2}\left(x_{1}^{2}+x_{2}^{2}\right)-E(x_{1},x_{2})\right)\nonumber \\
 & -J^{2},
\end{align}
solution of which, $E(x_{1},x_{2})$, is not a quadratic function
of vibrational mode coordinates $x_{1}$ and $x_{2}$.

In conclusion, by inspecting absorption spectrum of a wide range of
J and H molecular aggregates, in both CC and OC site configurations,
with various nearest neighbor coupling strength and temperature values,
we find the $\text{mD}_{2}$ Ansatz with depth of $M=7$ to be required
for accurate aggregate absorption spectra simulation, while the regular
Davydov $\text{D}_{2}$ Ansatz is not sufficient. For H aggregates,
multiplicity is required to obtain absorption lineshape positivity
and correct peak intensities. For J aggregates, increasing the number
of $\text{mD}_{2}$ Ansatz depth, mostly redshifts absorption spectrum,
keeping the overall lineshape qualitatively stays the same, especially
in CC aggregate. However, the very exchange narrowing effect is captured
by the simple Davydov $\text{D}_{2}$ Ansatz. Due to vibronic energy
level structure of an aggregate, we find molecular wavefunction to
exhibit an out-of-phase oscillatory behavior along the coordinate
and momentum axes and an overall broadening, which again is not captured
by the Davydov $\text{D}_{2}$ Ansatz.
\begin{acknowledgments}
We thank the Research Council of Lithuania for financial support (grant
No: SMIP-20-47). Computations were performed on resources at the High
Performance Computing Center, \textquoteleft \textquoteleft HPC Sauletekis\textquoteright \textquoteright{}
in Vilnius University Faculty of Physics.
\end{acknowledgments}

\appendix

\section{Multi-Davydov $\text{D}_{2}$ equations of motion and numerical implementation\label{sec:Appendix-A}}

Following the TDVP procedure, we derived vibronic molecular aggregate
EOMs, given by
\begin{align}
\sum_{j}\left(\dot{\alpha}_{j,n}S_{ij}+\alpha_{j,n}S_{ij}K_{ij}\right) & =-\text{i}\Theta_{i,n},\label{eq:eom1}
\end{align}
for each pair of indices $\left\{ i,n\right\} $, and
\begin{align}
\sum_{j,n}\left(\alpha_{i,n}^{\star}\dot{\alpha}_{j,n}S_{ij}\lambda_{j,kh}+P_{ij,n}\dot{\lambda}_{j,kh}\right)\nonumber \\
+\sum_{j,n}P_{ij,n}\lambda_{j,kh}K_{ij} & =-\text{i}\Omega_{i,kh},\label{eq:eom2}
\end{align}
for pair of $\left\{ i,k,h\right\} $ indices. These constitute a
system of equations needed to solve to propagate the $\text{mD}_{2}$
Ansatz, shown in Eq. (\ref{eq:mD2}). Dot notation is used, where
$\dot{x}$ is the time derivative of $x$. Right-hand side of given
EOMs are
\begin{align}
\Theta_{i,n} & =\sum_{j,m}\alpha_{j,m}S_{ij}J_{nm}\nonumber \\
 & +\sum_{j}\alpha_{j,n}S_{ij}\sum_{h}\left(C_{ij,nh}+\sum_{k}A_{ij,kh}\right),
\end{align}
\begin{align}
\Omega_{i,kh} & =\sum_{j,n,m}G_{ij,nm}\lambda_{j,kh}J_{nm}\nonumber \\
 & +\sum_{j,n}P_{ij,n}\lambda_{j,kh}\sum_{q}\left(C_{ij,nq}+\sum_{f}A_{ij,fq}\right)\nonumber \\
 & +\sum_{j}P_{ij,k}f_{kh}\omega_{kh}-\text{i}\sum_{j,n}P_{ij,n}\omega_{kh}\lambda_{j,kh},
\end{align}
where auxiliary definitions are
\begin{align}
G_{ij,nm} & =\alpha_{i,n}^{\star}\alpha_{j,m}S_{ij},\\
P_{ij,n} & =G_{ij,nn},\\
A_{ij,kh} & =\omega_{kh}\lambda_{i,kh}^{\star}\lambda_{j,kh},\\
C_{ij,nh} & =-f_{nh}\omega_{nh}\left(\lambda_{i,nh}^{\star}+\lambda_{j,nh}\right),\\
K_{ij} & =\sum_{m,h}\dot{\lambda}_{j,mh}\left(\lambda_{i,mh}^{\star}-\frac{1}{2}\lambda_{j,mh}^{\star}\right)\\
 & -\sum_{m,h}\frac{1}{2}\dot{\lambda}_{j,mh}^{\star}\lambda_{j,mh}.
\end{align}

We solved the presented system of EOMs in terms of variable $\alpha_{i,n}$,
$\lambda_{i,kh}$ real and imaginary parts, which are ordered in a
column state vector, $\boldsymbol{x}=\left\{ \boldsymbol{\alpha}_{i,n}^{\text{R}},\boldsymbol{\alpha}_{i,n}^{\text{I}},\boldsymbol{\lambda}_{i,kh}^{\text{R}},\boldsymbol{\lambda}_{i,kh}^{\text{I}}\right\} $.
This doubles the amount of variables, however, removes consistency
problems regarding treatment of complex variables $\dot{\lambda}_{j,mh}$,
$\dot{\lambda}_{j,mh}^{\star}$.

Numerical propagation of the $\text{mD}_{2}$ Ansatz is a two step
process. First, the time derivative of a state vector, $\dot{\boldsymbol{x}}$,
is found, by writing Eqs. (\ref{eq:eom1}), (\ref{eq:eom2}) in a
matrix form
\begin{equation}
\boldsymbol{M}\dot{\boldsymbol{x}}=\boldsymbol{f},\label{eq:Mxf}
\end{equation}
and solving for $\dot{\boldsymbol{x}}$ using the Generalized Minimal
Residual Method (GMRES) with Lower--Upper (LU) decomposition as a
preconditioner. We found GMRES method to provide a more accurate and
stable solution than using the Moore-Penrose pseudo inverse or the
solely LU decomposition method. Second, the state vector now can be
propagated using a variety of ordinary differential equation solvers
\citep{rackauckas2017}. We found an adaptive-order adaptive-time
Adams-Moulton method (VCABM) \citep{ErnstHairerGerhardWanner1993}
to provide just as accurate solution as a typical Runge--Kutta fourth-order
method, however, with less computational effort.

During time evolution of the $\text{mD}_{2}$ Ansatz, two, or more,
multiplicity wavepackets can approach each other and highly overlap,
this results in an ill-conditioned coefficient matrix, $\boldsymbol{M}$,
with no consistent solution of Eq. (\ref{eq:Mxf}). To remedy this,
we have implemented a programmed removal (apoptosis) of overlapping
multiples of the $\text{mD}_{2}$ Ansatz, with the minimal distance
for apoptosis to occur $d=0.05$, as defined in Ref. \citep{Werther2020}.

Establishing scaling factor of the numerical effort required to propagate
$\text{mD}_{2}$ Ansatz with increasing model size is not straightforward.
The total number of complex variables, $V$, describing $\text{mD}_{2}$
Ansatz is easy to find, $V=M\cdot\left(N+K\cdot Q\right)$, however,
due to first having to compute the time derivative of a state vector,
$\dot{\boldsymbol{x}}$, which involves non-linear and/or iterative
methods, the actual numerical effort is difficult to quantify. Empirical
estimation, which would compare scaling factors of several computation
approaches, is an interesting future research avenue.

\bibliographystyle{rsc}
\addcontentsline{toc}{section}{\refname}\bibliography{AGG}

\end{document}